\documentclass[preprint,prl,aps,amsfonts,amssymb,amsmath,showpacs]{revtex4}
\usepackage{graphicx}
\usepackage{dcolumn}

\graphicspath{{figures/}}
\setkeys{Gin}{width=\linewidth}

\begin{document}
\title{Imaging Coulomb Islands in a Quantum Hall Interferometer}

\author{B. Hackens$^{1}$, F. Martins$^{1}$, S. Faniel$^{1}$, C. A. Dutu$^{2}$, H. Sellier$^{3}$, S. Huant$^{3}$, M. Pala$^{4}$, 
L. Desplanque$^{5}$, X. Wallart$^{5}$, V. Bayot$^{1,3}$}
\affiliation{$^{1}$ Institute of Condensed Matter and Nanosciences - Nanophysics (IMCN/NAPS), 
Universit\'e catholique de Louvain, B-1348 Louvain-la-Neuve, Belgium\\
$^{2}$ Information and Communication Technologies, Electronics and Applied Mathematics (ICTEAM), 
Universit\'e catholique de Louvain, B-1348 Louvain-la-Neuve, Belgium\\
$^{3}$ Institut N\'eel, CNRS and Universit\'e Joseph Fourier, BP 166, 38042 Grenoble cedex 9, France\\
$^{4}$ IMEP-LAHC, Grenoble INP, Minatec, 3 Parvis Louis N\'eel, 38016 Grenoble, France\\
$^{5}$ IEMN, Cit\'e scientifique, Villeneuve d'Ascq, France\\ 
}

\date{\today}

\begin{abstract}
In the Quantum Hall regime, near integer filling factors,
electrons should only be transmitted through
 spatially-separated edge states. 
However, in mesoscopic systems, electronic transmission 
turns out to be more complex, giving rise to a large spectrum of magnetoresistance oscillations.
To explain these observations, recent models put forward that, as 
edge states come close to each other, 
electrons can hop between counterpropagating edge channels, 
or tunnel through Coulomb islands.
Here, we use scanning gate microscopy to demonstrate
 the presence of quantum Hall Coulomb islands, 
 and reveal the spatial structure of transport 
 inside a quantum Hall interferometer. 
 Electron islands locations are found by
modulating the tunneling between edge 
states and confined electron orbits. 
Tuning the magnetic field, we unveil 
a continuous evolution of active electron islands. 
This allows to decrypt the complexity of 
high magnetic field magnetoresistance oscillations, 
and opens the way to further 
local scale manipulations of quantum Hall localized states.
 
\end{abstract}

\maketitle

The quantum nature of electrons is two-sided: discreteness of charge and wave-like behaviour.
Most of the time, experiments reveal either one or the other. 
Quantum rings (QR) are known to be archetype devices 
where the wave-like nature of electrons manifests itself through Aharonov-Bohm 
(AB) oscillations\ \cite{AB_physrev59}.
They originate from the $B$-dependence of the phase 
acquired by electrons along the QR's arms, 
which modulates electron interferences within the QR. 
Therefore, in the coherent regime of transport and at low magnetic field $B$, 
the resistance of a mesoscopic quantum ring exhibits oscillations 
 as $B$ is varied, with a flux period corresponding 
 to the flux quantum $\phi_{0}=\mathrm{h/e}$\ \cite{Webb_PRL85}.
 
On the other hand, charge quantization can dramatically 
affect transport through a nanoscopic island occupied by $N$ electrons.
When the charging energy $\mathrm{e}^{2}/ C\gg \mathrm{k_{B}} T$, 
where $C$ is the island capacitance, Coulomb blockade (CB) prohibits 
electron transport except for potential values such that 
$N$ and $N+1$ states are degenerate\ \cite{Kouwenhoven_review}. 
CB manifests itself by periodic conductance peaks 
when the electrostatic potential is varied.

\begin{figure}[h]
\includegraphics[width=100 mm]{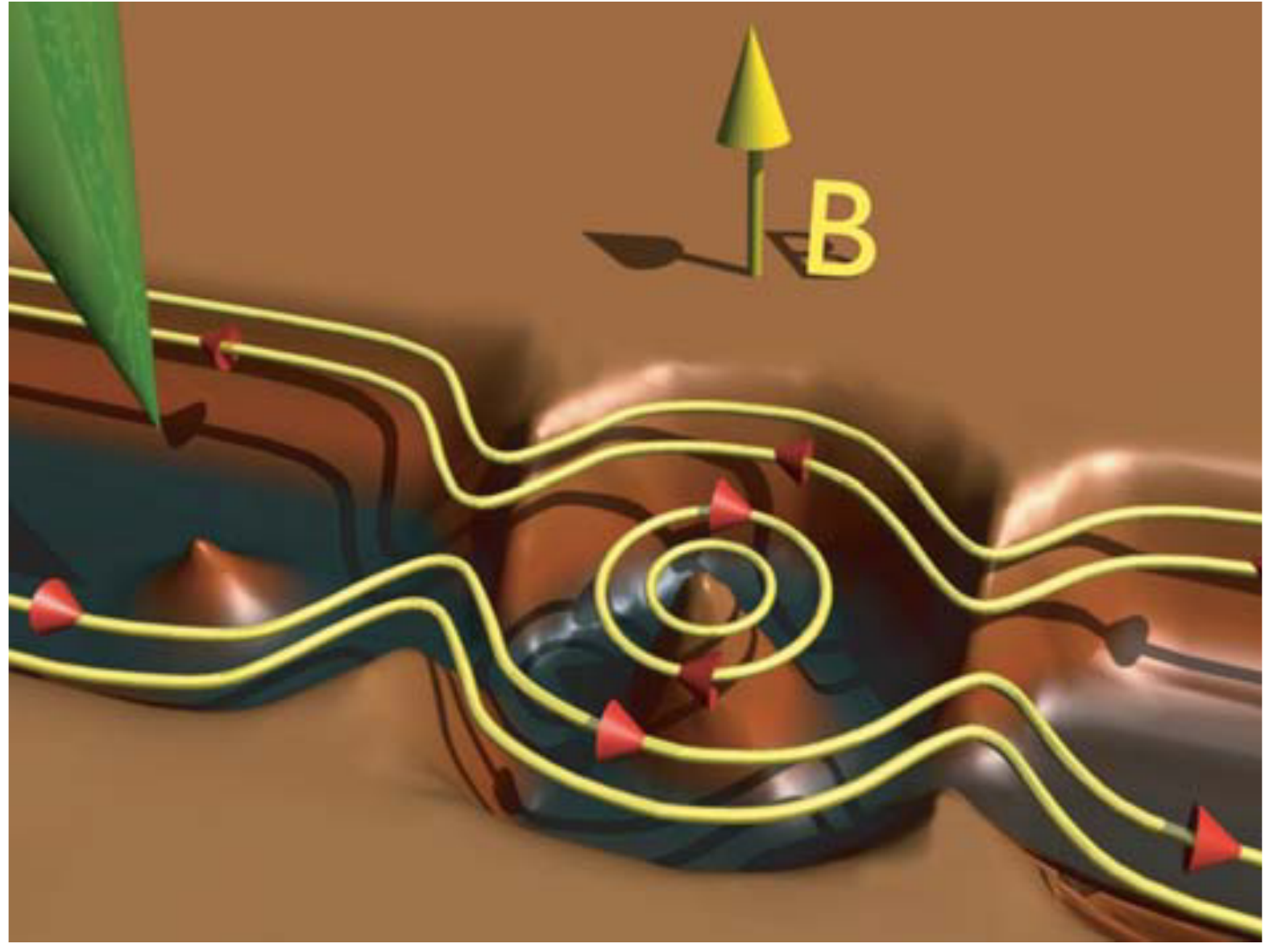}
\caption{{\bf Artist view of edge states in the quantum ring confining potential}. 
To simplify, we have assumed that no edge state is reflected at the ring's openings, 
which may not be the case in the real quantum ring.
The tip, in green, induces a local perturbation of the potential that 
can be scanned over the quantum Hall interferometer.}
\label{fig:fig0}
\end{figure}

By contrast, in the quantum Hall regime, the situation should be simpler : 
the resistance vanishes at low temperature since
transport occurs through edge states\ \cite{Halperin82,Buttiker88,chklovskiiPRB92}.
Several local-probe experiments have confirmed
this edge-state picture and imaged localized states
 in the bulk of macroscopic two-dimensional 
 electron systems\ \cite{Tessmer98,Yacobi99,Woodside01,hashimoto08}. 
However, numerous experiments on mesoscopic devices 
in the quantum Hall regime revealed 
 surprising observations, such as pseudo-Aharonov-Bohm "subperiods" 
 and "superperiods"\ \cite{GoldmanScience95,vanwees_PRL89,Taylor_PRL92,kataokaPRL99,CaminoPRB05,ZhangPRB09,giesbers10}. 
Motivated by these unexpected results, recent theories brought 
to the fore the discrete nature of electrons
to explain the broad range of periodicities. They invoke Coulomb interactions 
and tunneling between an electron island and edge states\ \cite{rosenowPRL07} (Fig.\ \ref{fig:fig0}), 
which may be combined with a spatial oscillation 
of the outermost electron orbit in the island\ \cite{giesbers10}. 
Complementing these theories, simulations yielded a detailed microscopic description 
of the different possible processes taking 
place in this regime\ \cite{IhnatsenkaPRB08,IhnatsenkaPRB09}.
To the theoretical debate adds the diversity of experimental observations, 
 which hampers unambiguous determination of the precise mechanisms 
at play at the local scale.

Here, we present a spatially-resolved investigation of 
electron transport inside an interferometer formed by an InGaAs/InAlAs 
quantum ring (see method section),
driven in the integer quantum Hall (QH) regime. 
We show that each pseudo-AB period can be associated with a specific Coulomb island 
formed by edge states loops enclosing a hill or a valley in the potential.
Each active Coulomb island can be located precisely inside the QR by tuning the magnetic field
and imaging the spatial shift of Coulomb resonances by means of scanning gate microscopy.

\section*{Results}
\begin{figure}[h]
\includegraphics[width=80 mm]{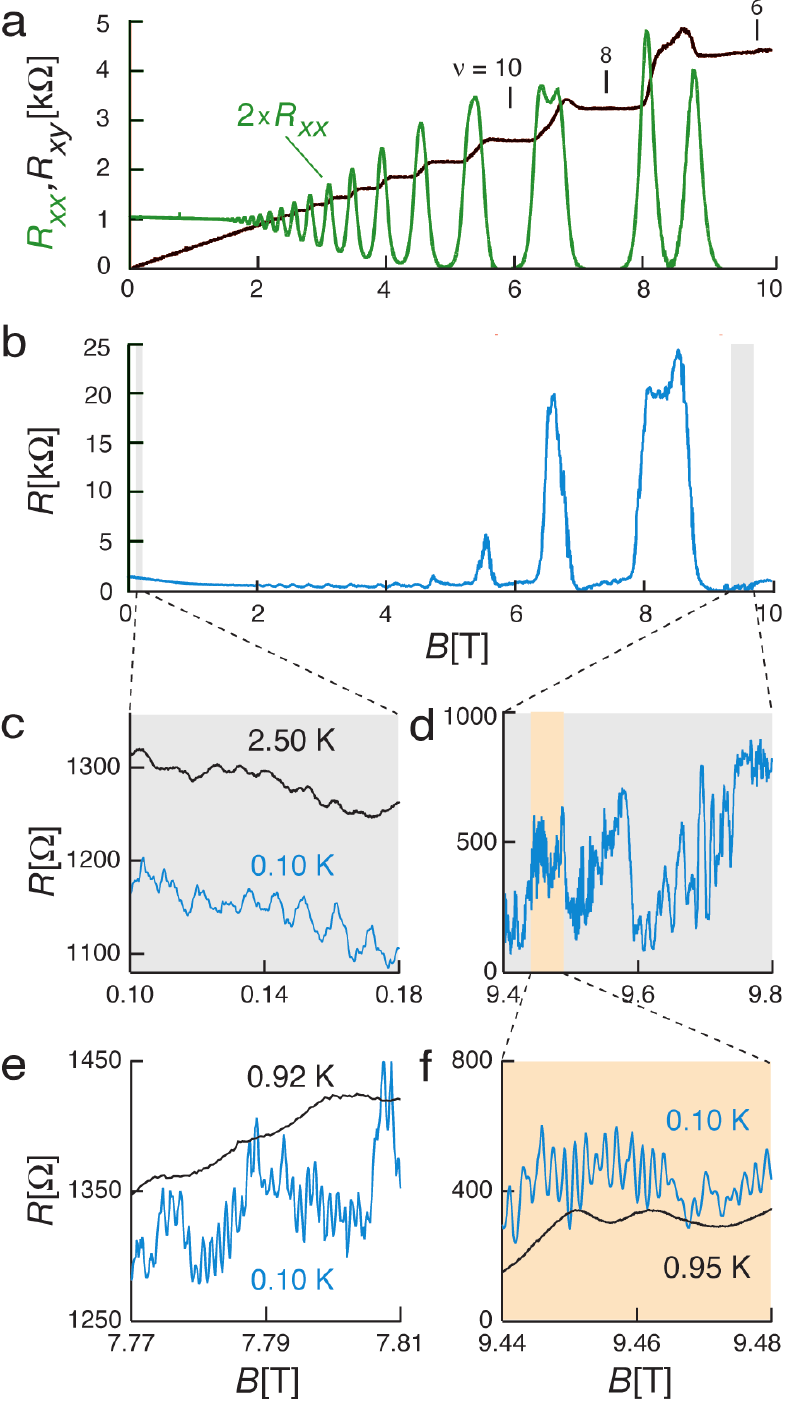}
\caption{{\bf Magnetoresistance measurements on the Hall bar and on the quantum ring.}
(a) Longitudinal and transverse (Hall) resistances $R_{xx}$ and $R_{xy}$ vs $B$, measured in the Hall bar.
Overshoots in $R_{xy}$ originate from a mixing of sthe longitudinal and Hall resistances, due to 
a geometrical asymmetry of the Hall bar. 
(b) Magnetoresistance of the QR. 
(c) Low-$B$ magnetoresistance of the QR, measured at $T=0.1\ \mathrm{K}$ and 2.5 K. The 
periodicity of the oscillations in $R$ vs $B$ is $\Delta B = 9$ mT.
(d) Magnetoresistance of the QR on the QH plateau around $\nu=6$, at $T=0.1\ \mathrm{K}$.
(e) Magnetoresistance of the QR on the QH plateau around $\nu=8$, $T=0.1\ \mathrm{K}$ and 0.92 K.
The periodicity of the oscillations in $R$ vs $B$ is $\Delta B = 1.1$ mT.
(f) Close-up view of Fig.\ \ref{fig:fig1}d in the orange-shaded region, at $T=0.1\ \mathrm{K}$ and 0.95 K. 
The periodicity of the oscillations in $R$ vs $B$ is $\Delta B = 1.5$ mT.
}
\label{fig:fig1}
\end{figure}

{\bf Magnetoresistance measurements.} At low $B$ and $T$, electron transport through the QR is in the coherent regime.
This is attested by periodic Aharonov-Bohm oscillations in resistance $R$ vs $B$.
As expected, the AB period $\Delta B_{\mathrm{AB}}= 9\ \mathrm{mT}$ (Fig.\ \ref{fig:fig1}c), 
corresponds to  the average QR radius (380 nm). 
Note also that the QR is in the quasi-ballistic regime 
(the electron mean free path is $\sim 800$ nm at low $B$), and $\sim$ 
25 transverse modes are transmitted through the QR's openings 
(the Fermi wavelength is 22 nm, and the openings' width is 250 nm).
At high-$B$, the two-dimensional electron system enters into 
the QH regime and its transverse magnetoresistance 
$R_{xy}$ vs $B$ exhibits plateaus at values $\mathrm{h}/(\nu \mathrm{e}^{2})$,
 where $\nu$ is the (integer) filling factor (Fig.\ \ref{fig:fig1}a). 
Within the plateaus, the Hall bar longitudinal magnetoresistance 
$R_{xx}$ vs $B$ drops to zero.
In contrast, in the $B$-range coinciding with the position 
of the plateaus, the magnetoresistance of the QR
displays strong reproducible fluctuations with a 
wide range of characteristic $B$-scales (Fig.\ \ref{fig:fig1}b-f).
On Fig.\ \ref{fig:fig1}b, we also 
note the absence of finite resistance plateaus, which would indicate 
a difference in the number of filled Landau levels in the bulk and 
in the constrictions\ \cite{CaminoPRB05}. 
A closer look at $R$ vs $B$ in the region near $\nu=6$ 
reveals high-frequency periodic oscillations over 
some parts of the high $B$-range (Fig.\ \ref{fig:fig1}f), 
with a period $\Delta B= 1.5\ \mathrm{mT}$ much smaller than the
 AB oscillations. Fig.\ \ref{fig:fig1}e shows that a similar pattern is also visible on the plateau near $\nu=8$, 
but with an even shorter period $\Delta B= 1.1\ \mathrm{mT}$. 
Finally, these high-$B$ oscillations rapidly 
decay with temperature (Fig.\ \ref{fig:fig1}e-f) and vanish before reaching 1 K.
This contrasts with AB oscillations that, like the phase coherence time, 
saturate below $\sim$1 K\ \cite{hackensPRL05} and are still visible at 2.5 K (Fig.\ \ref{fig:fig1}c). 
An alternative mechanism, unrelated to electron interferences,
must therefore be invoked to explain our observations.

To explain the presence of subperiod oscillations, a
recent theory invokes Coulomb blockade 
of electrons tunneling between the conducting edge states
transmitted along the borders of the 
QR and those forming a quantum Hall electron
island located at the center of the device\  
\cite{rosenowPRL07} (Fig.\ \ref{fig:fig0}).
Changing the magnetic field redistributes
electrons within the Landau levels (LLs ): if the flux through the area $A$
 enclosing the island increases by one flux quantum $\phi_{0}$, 
one electron must be added to each filled LLs. 
This process introduces a new contribution to the island charging energy,
with a magnetic field period $\Delta B=(\phi_{0}/A)/f_{c}=\Delta B_{\mathrm{AB}}/f_{c}$, 
where $f_{c}$ is indeed the number of filled LLs. 
In turn, charging energy oscillations affect the QR resistance through 
the Coulomb blockade of electrons tunnelling between 
the quantum Hall island and the transmitted edge states.
Note that the area encircled by edge states
 at high $B$ is not necessarily equal to the mean 
 area $A$ of the QR, which determines $\Delta B_{\mathrm{AB}}$. 
Simulations (not shown here) of the electron local density of states at 
$B=9.5$ T in a 2DEG with the same Fermi energy 
as in the experiment, indicate that the outermost populated edge state extends 
55 nm away from the edge of the device. 
Taking this value and edge roughness into account, an edge state 
loop encircling the central QR antidot would therefore have a radius 
similar to the QR mean radius. 
This explains the remarkable consistency between the prediction 
of pseudo-AB subperiods $\Delta B_{\mathrm{AB}}/f_{c}$ 
and our observations $\Delta B =  1.5$\ mT and 1.1\ mT
 at $\nu=6$ and $\nu=8$, respectively (Fig.\ \ref{fig:fig1}e-f).

From the analysis of the QR magnetoresistance oscillations, a
 quantum Hall Coulomb island should therefore
be present in our quantum ring. 
While our sample geometry is \emph{a priori} well suited to generate 
a quantum island around the central antidot, it could, as we will see later, 
locate elsewhere. Therefore local-scale information is needed to strenghten our understanding. 
In addition, the observation of subperiod oscillations only in narrow $B$-windows 
 remains to be explained.

{\bf Scanning gate microscopy.} Combining electron transport measurements with
local probe techniques has already proven
powerful to investigate the local details of electron behaviour 
inside various mesoscopic systems such as quantum point 
contacts\ \cite{Topinka:2000uq,Topinka:2001lr,aoki:155327}, 
open and closed quantum dots\ \cite{Woodside:2002dq,piodaPRL04,crook:246803}
and quantum interferometers in the Aharonov-Bohm regime  
\ \cite{Hackens:2006qf,Hackens_PRL07,PalaPRB08,PalaNano09}.
The technique, namely the scanning gate microscopy (SGM),
consists in scanning the electrically-biased tip of an 
atomic force microscope in a plane parallel to the 2DES and 
recording the changes in the QR resistance, $R(x,y)$, 
induced by the tip located at $(x,y)$. 
In the context of the present experiments,
the intrinsic sensitivity of Coulomb blockade to potential variations, 
makes SGM potentially well-suited to explore Coulomb islands.

\begin{figure}[h]
\includegraphics[width=100 mm]{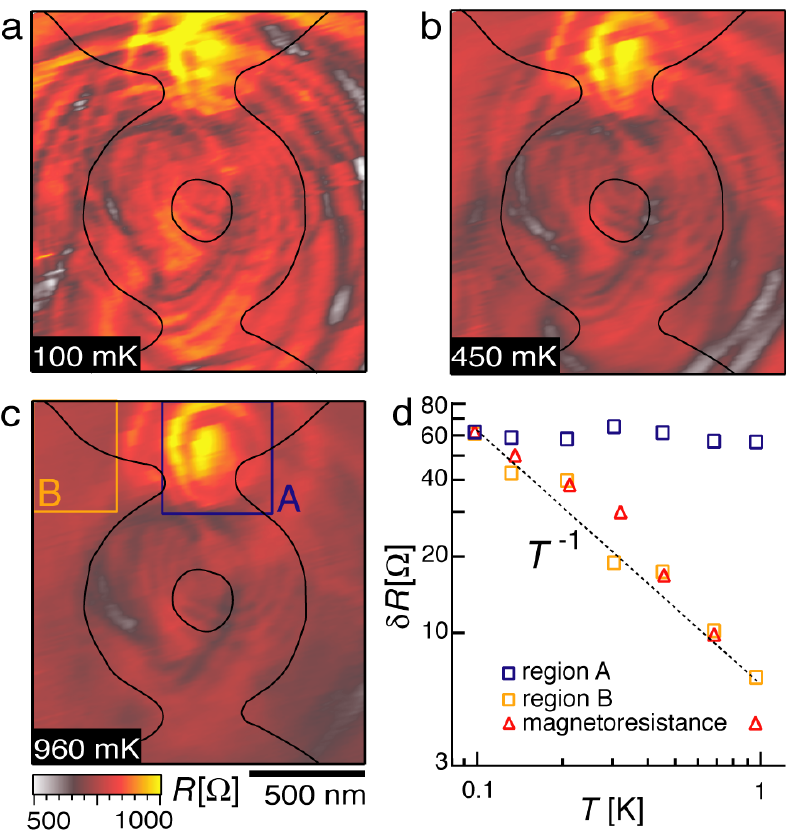}
\caption{{\bf Scanning gate resistance maps of the quantum ring, 
showing Coulomb islands-related concentric rings}.
 (a-c) $R(x,y)$ map obtained 
at $B=9.5\ \mathrm{T}$, $V_{\mathrm{tip}}=-1$ V
and at $T=$ 100, 450 and 960 mK, respectively. 
The black curves correspond to the position of the QR. 
(d) $T$-dependence of $\delta R$, calculated on raw $R(x,y)$ maps in regions A and B
 defined on Fig.\ \ref{fig:fig2}c (blue and orange squares, respectively)
and on high-pass filtered 
$R$ vs $B$ traces (red triangles).
The frequency of the high-pass 
 filter on $R$ vs $B$ curves is 60 T$^{-1}$, and $\delta R$ in this case
 is evaluated in the range 9.47 to 9.53\ T.}
\label{fig:fig2}
\end{figure}

In the $B$-range where subperiod oscillations were evidenced, 
the resistance map $R(x,y)$ reveals a first set of
concentric fringes (type-I) centered on the QR antidot
(Fig.\ \ref{fig:fig2}a). Qualitatively, this is exactly 
the behaviour expected for a gate scanning above an electron 
island experiencing Coulomb blockade, \emph{i.e.}
fringes correspond to isopotential lines located at 
constant distances from the island\ \cite{Woodside:2002dq,piodaPRL04}.
It is worth noting that it is indeed the flux through 
the edge state loop that controls the present   
CB effect. Hence, it can be tuned either by sweeping $B$ 
or by approaching a negatively-polarized tip 
which raises the potential and enlarges the loop surface.
Each concentric type-I fringe in Fig.\ \ref{fig:fig2}a can 
therefore also be viewed as an \emph{isoflux} line, 
that marks one single CB state.
Note that the observed succession of concentric fringes is reminiscent 
of patterns evidenced in previous SGM experiments 
on a QR at low-$B$\ \cite{Hackens:2006qf}, but the physical origin is totally different. 
As they vanish with increasing $T$, 
type-I fringes leave more apparent a second set of fringes (type-II), 
which is found $T$-independent below 1 K (Fig.\ \ref{fig:fig2}b-c).
The presence of two distinct phenomena
is made even clearer on Fig.\ \ref{fig:fig2}d,
representing the $T$-dependence of $\delta R$, 
the standard deviation of $R(x,y)$ in 
the rectangular regions labeled A and B on Fig.\ \ref{fig:fig2}c, 
which are mainly encompassing type-I and type-II fringes, respectively. 

\section*{Discussion}
Focusing on type-I fringes we note that they exhibit 
the same $T$ dependence as the subperiod oscillations 
in the magnetoresistance (Fig.\ \ref{fig:fig2}d).
More precisely, the $T^{-1}$ dependence of $\delta R$ 
is compatible with a Coulomb blockade (CB) origin for both observations.
Indeed, in classical CB experiments at $B = 0$ T, 
the conductance peak height was reported to scale as
 $\delta G \propto T^{-1}$\ \cite{Folk_PRL96,Kouwenhoven_review}, 
 in the case of single-level transport through 
 a lateral quantum dot. In our case, electrons transmitted through 
Coulomb islands contribute only to a part of the total 
conductance ($\delta G/G\approx\delta R/R  \lesssim 0.1 $). 
Therefore, $\delta R$ and $\delta G$ should effectively exhibit 
the same $T$ dependence.

Different frameworks can be invoked to explain the origin of type-II fringes. 
First, in the QH regime, potential variations are at the origin of 
'hot spots', \emph{i.e.} regions of high sensitivity of 
the resistance to local potential changes, which have been evidenced  
by SGM in a Hall bar at 1.9 K\ \cite{baumgartner:085316}.
This may explain type-II patterns observed \emph{within the area of the QR}.  
But the concentric ring-like patterns at the QR opening (region A, Fig.\ \ref{fig:fig2}c)
cannot be explained within this model, as 
'hot spots' do not give rise to an oscillatory behaviour of the resistance
when the tip approaches the spot.
As an alternative, we suggest that the tip can modulate 
tunneling processes between counterpropagating 
edge states\ \cite{rosenowPRL07} at the openings. Affecting more
 than one edge state with the tip potential would 
 induce the observed concentric pattern. 
Magnetoresistance measurements could confirm this scenario : 
indeed, $R$ vs $B$ should exhibit finite resistance 
plateaus in the case of a tip-induced difference in the number of 
filled Landau levels inside the QR and at the openings\ \cite{CaminoPRB05}.

\begin{figure}[h]
\includegraphics[width=100 mm]{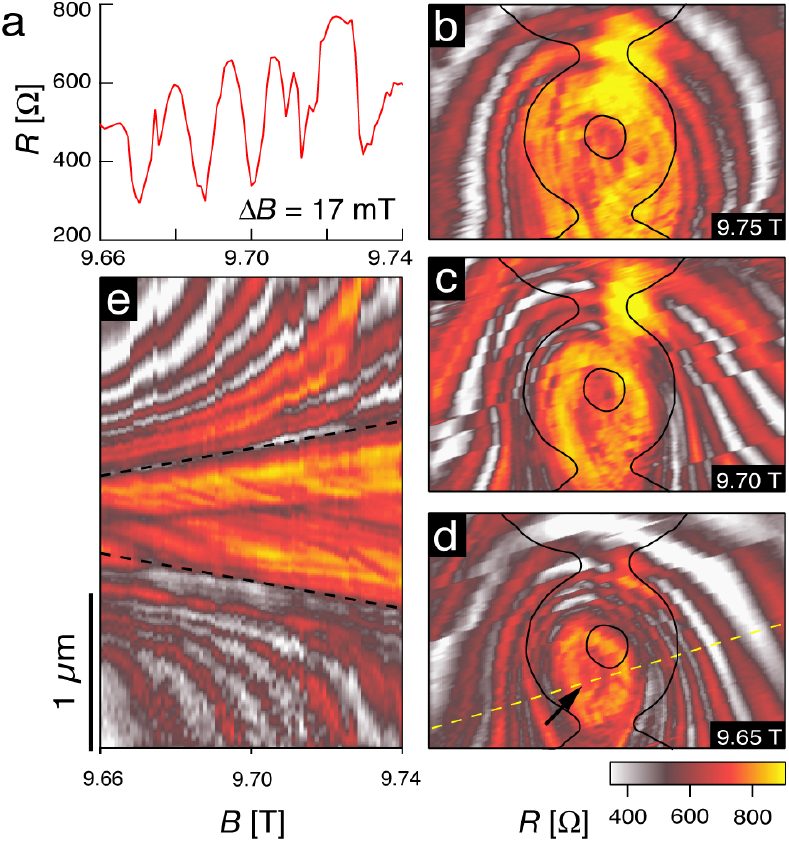}
\caption{{\bf Changing the magnetic fields reveals the position 
of a Coulomb island on scanning gate microscopy maps.} (a) $R$ vs $B$ at $T=100$ mK. 
(b-d) $R(x,y)$ maps obtained at $B=$ 9.75, 9.70 and 9.65 T, respectively, with 
$T=100$ mK and $V_{\mathrm{tip}}$=-1 V. (e) $B$-dependence of the $R(x,y)$ profile 
measured along the dashed line in Fig.\ \ref{fig:fig3}d.}
\label{fig:fig3}
\end{figure}

Up to now, we have only analyzed a small $B$-range 
where $\Delta B_{AB}/f_{c}$ subperiods dominate.
A close look at Fig.\ \ref{fig:fig1}d reveals that 
changing the magnetic field only by a few 
percents (\emph{e.g.} from 9.5 to 9.7\ T) 
dramatically modifies the magnetoresistance.
This time, marked AB-like 'superperiod' 
oscillations show up, with $\Delta B=$17\ mT (Fig.\ \ref{fig:fig3}a), 
which share the same $T^{-1}$ dependence (not shown) as the subperiods discussed above.
According to the Rosenow-Halperin model\ \cite{rosenowPRL07}, 
the superperiod oscillations indicate much smaller edge state 
loops. With an enclosed area $A=(\phi_{0}/\Delta B)/f_{c}=4 \times 10^{-14}\ \mathrm{m}^{2}$, 
the loop should have a mean radius of $\sim$ 65\ nm, 
and hence be located somewhere into the 300 nm-wide QR's arms. 
Our scanning gate should therefore be able to locate the loop. 

The SGM localization is achieved in the sequence of Fig.\ \ref{fig:fig3}b-d. 
From 9.75 to 9.65 T, fringes shrink and finally 
reveal the active Coulomb island, 
marked by the arrow on Fig.\ \ref{fig:fig3}b. 
Note that the precise shape of the central 
SGM pattern is a convolution of the Coulomb island and tip potential shapes\ \cite{Gildemeister_PRB07}.
A more detailed view of this process is shown 
in Fig.\ \ref{fig:fig3}e, displaying
the $B$-dependence of the SGM profile taken 
along the dashed line in Fig.\ \ref{fig:fig3}d, 
crossing the center of the fringes. Clearly, isoresistance lines shift towards lower $B$ as the negatively-charged tip approaches.
Interestingly, the sign of the $B$-shift can indeed discriminate 
between edge states surrounding a potential hill 
and those confined on the border of the potential well.
Assuming that the incompressible island surrounds a potential hill,
approaching a negative-biased 
tip will raise the potential and increase the loop surface. 
If the incompressible island were localized in a potential well, 
the tip would reduce the loop surface.
Since isoresistance lines correspond to isoflux states through the incompressible loop,
the low-$B$ shift in Fig.\ \ref{fig:fig3}e unambiguously 
identifies that the loop surrounds a potential hill.

It is worth noting that both subperiods 
and superperiods are observed in limited $B$-ranges.
This means that a small potential perturbation by the tip could 
completely wipe out the CB oscillations, 
by either suppressing tunneling channels or suppressing CB islands.
The region delimited by the dashed lines in Fig.\ \ref{fig:fig3}e 
provides an example of such a 
state where AB-like superperiods are suppressed.

\begin{figure}[h]
\includegraphics[width=100 mm]{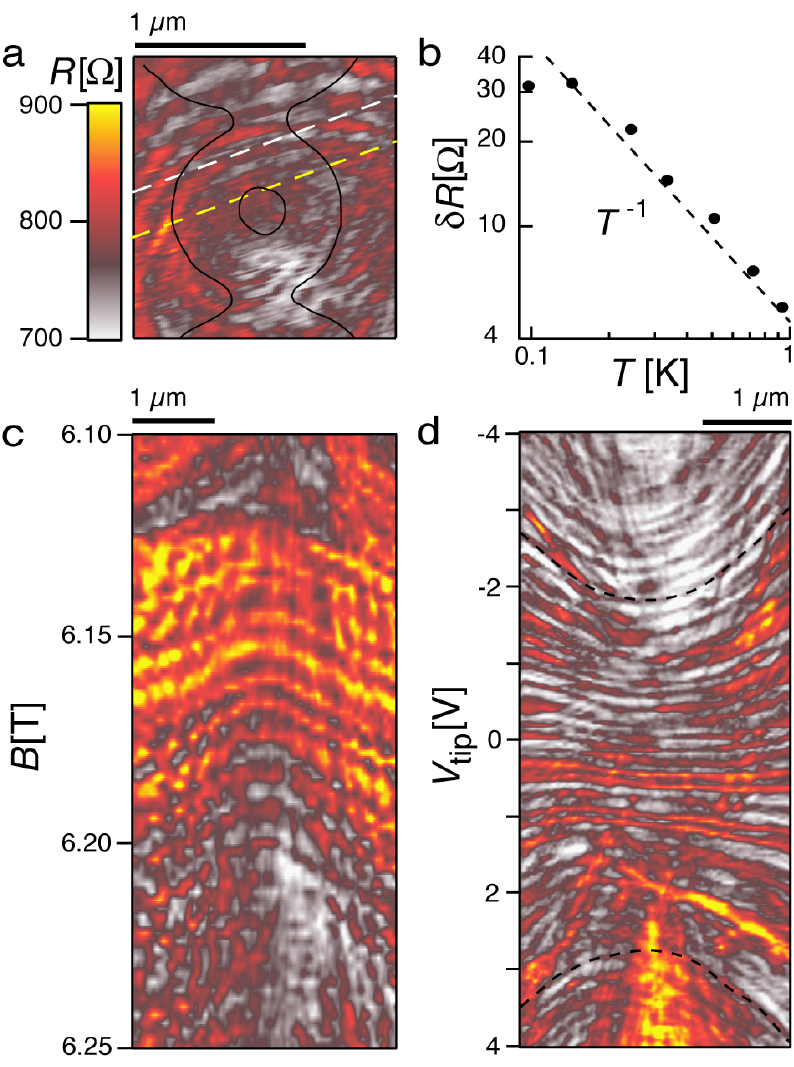}
\caption{{\bf Shift of the Coulomb blockade resonances with the magnetic field and 
tip voltage.} (a) $R(x,y)$ map measured at $B=6.25$ T and $V_{\mathrm{tip}}=-0.5$ V. 
(b) $T$-dependence of $\delta R$ measured from a $R(x,y)$ map at $B=6.25$\ T. 
(c) $B$-dependence of the $R(x,y)$ profile 
measured along the yellow dashed line in Fig.\ \ref{fig:fig4}a.
(d) $V_{\mathrm{tip}}$-dependence of the $R(x,y)$ profile measured 
along the white dashed line in Fig.\ \ref{fig:fig4}a. The black line at the top of each graph represents 1\ $\mu$m.}
\label{fig:fig4}
\end{figure}

At first sight, the data taken near $\nu=10$ and shown in 
Fig.\ \ref{fig:fig4}a resembles more SGM maps presented in 
another QR but at low B \cite{Hackens:2006qf}. 
Strikingly, the $B$-dependence of the profile taken 
along the yellow dashed line in Fig.\ \ref{fig:fig4}a 
shows dominant oscillations with $\Delta B\sim$ 7.8 mT (Fig.\ \ref{fig:fig4}c), 
very close to what the orthodox AB effect would provide 
in our sample. But the temperature dependence, 
shown in Fig.\ \ref{fig:fig4}b, is not that expected for the AB effect. 
Indeed, the observed resistance oscillations are only 
pseudo-AB ones as they exhibit the same $T^{-1}$-dependence 
as sub and superperiod oscillations, 
indicating that they share the same Coulomb blockade origin.

As already shown near $\nu=6$, the CB resonances 
shift towards low-$B$ when the tip 
approaches the QR (Fig.\ \ref{fig:fig4}c). 
However, the relative amplitude of this shift is smaller compared
 to Fig.\ \ref{fig:fig3}e because the tip does not scan directly over the active
Coulomb island, so the potential perturbation 
experienced by the Coulomb island is not as large as in Fig.\ \ref{fig:fig3}e.
In addition, the data reveal a much richer pattern of fringes 
indicating that multiple Coulomb islands are at work.
This is confirmed in Fig.\ \ref{fig:fig4}d that shows the $V_{tip}$ dependence 
of SGM profiles taken along the white dashed line on Fig.\ \ref{fig:fig4}a 
(note that changing the position of the SGM profile yields qualitatively similar results). 
The main features are parabolic-like iso-resistance lines, indicated by the dashed lines. 
Their curvature evolves from zero at $V_{tip}=0$ V 
to positive or negative, according to the polarity of the tip, 
as expected for Coulomb interaction between the tip and 
a potential hill (see also ref.\ \cite{piodaPRL04}). However, 
Fig.\ \ref{fig:fig4}d, like Fig.\ \ref{fig:fig4}c, also 
shows steeper resistance lines confirming the richness 
and complexity of mesoscopic transport in quantum Hall interferometers.

The wealth of behaviours exhibited in SGM data 
at different magnetic fields and tip voltages illustrates
the capability of SGM to embrace the complexity 
of quantum Hall interferometers at the local scale, 
and, beyond that, of mesoscopic transport. 
The fine tuning of the local scale 
potential by the biased tip offers various 
opportunities to image and manipulate localized electron states. 
Such a detailed probing of electron transport 
in quantum Hall interferometers
may also prove valuable in future explorations
of non abelian quantum states\ \cite{dassarmaPRL05,Stern_PRL06}.

\section*{Methods}
{\bf Device fabrication and sample parameters.} The quantum ring was
patterned using electron beam lithography and 
wet etching in an InGaAs/InAlAs heterostructure\ \cite{HackensPRB02}. 
A two-dimensional electron system is confined
 25 nm below the surface and its low temperature
 electron density and mobility are $1.4 \times 10^{16}\ \mathrm{/m^{2}}$
and $4\ \mathrm{m^{2}/Vs}$, respectively, calculated from longitudinal and transverse 
magnetoresistance ($R_{xx}$ and $R_{xy}$ vs $B$, Fig.\ \ref{fig:fig1}a)
measurements on a Hall bar patterned next to the QR.
The quantum ring inner and outer lithographic radii are 215 nm and 520 nm, 
respectively, and the width of both openings is 300 nm (Fig. S2).
Note that the depletion length at the edge of etched trenches is $\sim 25\ \mathrm{nm}$. 

{\bf SGM technique.} 
The experiments are carried out down to 100 mK with the sample thermally 
anchored to the mixing chamber of a $^{3}\mathrm{He}/^{4}\mathrm{He}$ 
 dilution refrigerator, and a home-made 
Atomic Force Microscope (AFM) suspended underneath. 
The force sensor consists in a commercial AFM cantilever 
coated with 20 nm of Ti and 10 nm of Pt
(model CSC17 from MikroMasch; nominal tip radius : 40 nm), glued on 
one prong of a quartz tuning fork 
using conductive silver epoxy (Fig. S1a). 
To measure the sample topography, the AFM is operated 
in the dynamic mode with a feedback loop on the 
tuning fork oscillation frequency shift. 

Using this method, we obtained the topography image of 
the quantum ring shown in Fig. S2a, at a temperature $T=100$ mK. 
Clear edges are visible where the heterostructure was etched, 
which is an indication of a sharp tip.
Further information on the quality of both the tip and 
the imaging technique can be inferred
 by comparing the topography image (Fig. S2a) with Fig. S2b, 
obtained by Scanning Electron Microscopy (SEM) before cooling down the sample.
Small differences can be observed between Fig. S2a and b.
In particular, the size of the etched trenches is $\sim 100$ nm smaller in the AFM topography.
This discrepancy originates from the convolution of the tip shape 
with the surface topography in the AFM picture.
Fig. S1b shows an image
of the extremity of the tip used in the experiments presented in this work,
obtained by SEM after the end of the SGM experiments. 
 The size of the tip apex on this picture is $\sim 70$ nm,
which explains the difference in dimensions between 
Fig. S2a and b.

After imaging the topography of the ring, the tip is lifted at a distance 
of $\sim$50 nm from the sample surface (\emph{i.e.} 
75 nm away from the 2D electron system), and
a bias voltage $V_{\mathrm{tip}}$ is applied on the tip 
in order to induce a local electrostatic 
 perturbation for electrons transmitted through the QR.
The quantum ring resistance $R$ is
measured using a low frequency (28 Hz) lock-in technique
 with a source-drain voltage across the QR always less than k$T$/e.
Figures 4b-d were measured over several hours.
Before and after such a set of SGM experiment, the QR topography was imaged in 
order to check if the tip drifted with respect to the QR position during the experiment. 
We observed that the lateral drift was always smaller than 50 nm.

An additional shift between the topography and the $R(x,y)$ map
 may originate from differences between the topographic tip 
and the "electric tip", \emph{i.e.} the perturbation of
 electrostatic potential associated with
the charged tip, experienced by electrons inside the quantum ring. 
Indeed, charged contamination on both the tip or the 
sample surface may contribute to alter this potential.
This shift can be roughly estimated 
by noting that each set of concentric resistance fringes 
in the $R(x,y)$ maps is centered on a Coulomb island,
and must therefore have its center located within the limits of the 
quantum ring. By examining all the available $R(x,y)$ maps, we could 
determine a higher bound of $\sim$ 150 nm on this shift 
(\emph{i.e.} a shift larger 
than 150 nm would mean that some sets of resistance fringes 
are centered outside the QR area).

A crucial issue in the context of scanning gate microscopy is 
the "gating behaviour" of the tip \cite{Gildemeister_PRB07,Gildemeister_APL07}. 
First, Fig. S1b shows that the metal layer
covering the tip apex remained continuous during our experiment, 
and was not damaged by the numerous topographic scans realized
 at low temperature. In addition, there is no indication for a multiple tip,
 neither in Fig. S1b, nor in topography mode, nor in SGM images.
Complementary, the electrical quality of the tip can be evaluated from the $R(x,y)$ maps. 
In the $R(x,y)$ map shown in Fig. S3, neighboring resistance fringes
separated by less than $\sim 50$ nm can be distinguished, 
which gives an indication on the width of the perturbating potential.
Note that this width increases with the tip-surface separation in SGM mode.

Furthermore, the electrical potential associated with the charged tip 
should ideally have a circular symmetry, in order to avoid unwanted distortion of the features
visible in the $R(x,y)$ map. This symmetry can be evaluated 
on SGM maps of quantum dots: when the tip potential shifts an energy level
of the quantum dot in resonance with the source and drain energy levels, electrons
can tunnel through the quantum dot, which induces a current rise.
In an SGM map, these resonances are visible as sets of concentric rings. 
The set of rings is directly related to the electrical potential of the 
tip, as shown in ref.\ \cite{Gildemeister_PRB07}. In our experiment, 
the sets of resistance fringes in $R(x,y)$ maps are also 
related to the Coulomb blockade phenomenon, 
and correspond to isopotential lines. Their precise shape
depends both on the tip potential shape, and 
on the shape of the potential hill around 
which each electron loop is localized.
In the case of an anisotropic tip potential, 
the sets of resistance fringes in $R(x,y)$ maps would all exhibit the same 
shape distortion related to the potential anisotropy.
This is contrary to our observations: we observe a large diversity
in the shapes of resistance fringes, but no recurrent anisotropy. 
This indicates that the tip potential shape is not anisotropic.

\section*{Acknowledgements}
B. H. is postdoctoral researcher with the Belgian FRS-FNRS 
and F. M. is funded by FNRS and FCT (Portugal) postdoctoral grants. 
This work has been supported by FRFC grant no. 2.4.546.08.F,
and FNRS grant no 1.5.044.07.F,
by the Belgian Science Policy (Interuniversity Attraction Pole Program IAP-6/42),
This work has also been supported by the PNANO 2007 program of
the Agence Nationale de la Recherche (`MICATEC' project).
VB acknowledges the award of a `Chaire d'excellence'
by the Nanoscience Foundation in Grenoble.

\section*{Author contributions}
F. M. and B. H. performed the low-temperature SGM experiment; B. H., F. M. and V. B. analyzed the experimental data; L. D. and X. W. grew the InGaAs heterostructure; B. H. and C. A. D. processed the samples; B. H., S. F. and F. M. built the low temperature scanning gate microscope; B. H., F. M., S. F., H. S., S. H., M. P. and V. B. contributed to the conception of the experiment; B. H. and V. B. wrote the paper and B. H., F. M., H. S., S. H., M. P. and V. B. discussed the results and commented on the manuscript.

\section*{Competing financial interests}
The authors declare no competing financial interests.

\newpage

\end{document}